\PassOptionsToPackage{table,xcdraw}{xcolor}
\documentclass[12pt,a4paper]{article}
\pdfoutput=1

\usepackage{geometry}
\usepackage[numbers,sort&compress]{natbib}
\usepackage{amsmath}
\usepackage{amssymb}
\usepackage[dvips]{graphicx}
\usepackage{pstricks}
\usepackage{bm}
\usepackage{pbox}
\usepackage{placeins}
\usepackage{graphicx}
\usepackage{caption}
\usepackage{subcaption}
\usepackage[T1]{fontenc}
\usepackage[utf8]{inputenc}
\usepackage{footnote}
\usepackage{pdfpages}
\usepackage{hhline}
\usepackage{multirow}
\usepackage{enumitem}
\usepackage[nottoc,notlot,notlof]{tocbibind}
\usepackage[title,titletoc]{appendix}
\usepackage{dsfont}
\allowdisplaybreaks
\usepackage{cases}

\definecolor{MyDarkBlue}{rgb}{0.1, 0.1, 0.8} 
\definecolor{MyLightBlue}{rgb}{0.22,0.51,0.9}
\definecolor{MyGreen}{rgb}{0.0, 0.5, 0.0}
\definecolor{BrickRed}{rgb}{0.8, 0.25, 0.33}
\usepackage[colorlinks=true, linkcolor=blue, citecolor=MyDarkBlue, urlcolor=MyLightBlue, bookmarksnumbered=true, bookmarksopen]{hyperref}
\hypersetup{colorlinks, citecolor=blue,linkcolor=black,  urlcolor=MyLightBlue}

\begin{document}
\vspace*{-0.2in}
\begin{flushright}
OSU-HEP-20-06
\end{flushright}
\vspace{0.5cm}

\begin{center}
{\Large\bf 
$\mu \to e \gamma$ selecting scalar leptoquark solutions for the $(g-2)_{e,\mu}$ puzzles}\\
\end{center}

\vspace{0.5cm}
\renewcommand{\thefootnote}{\fnsymbol{footnote}}
\begin{center}
{\large
{}~\textbf{Ilja Dor\v{s}ner}$^{a,b,}$\footnote{ E-mail: \textcolor{MyLightBlue}{dorsner@fesb.hr}}, 
{}~\textbf{Svjetlana Fajfer}$^{b,c,}$\footnote{ E-mail: \textcolor{MyLightBlue}{svjetlana.fajfer@ijs.si}}, and 
{}~\textbf{Shaikh Saad}$^{d,}$\footnote{ E-mail: \textcolor{MyLightBlue}{shaikh.saad@okstate.edu}}
}
\vspace{0.5cm}

$^{a}${\em  University of Split, Faculty of Electrical Engineering, Mechanical Engineering and Naval Architecture in Split (FESB), Ru\dj era Bo\v{s}kovi\'{c}a 32, 21000 Split, Croatia}\\
$^{b}${\em  Jo\v{z}ef Stefan Institute, Jamova 39, P.\ O.\ Box 3000, 1001 Ljubljana, Slovenia}\\
$^{c}${\em Department of Physics, University of Ljubljana, Jadranska 19, 1000 Ljubljana, Slovenia}\\
$^{d}${\em Department of Physics, Oklahoma State University, Stillwater, OK 74078, USA }
\end{center}

\renewcommand{\thefootnote}{\arabic{footnote}}
\setcounter{footnote}{0}
\thispagestyle{empty}

\begin{abstract}
We investigate all potentially viable scenarios that can produce the chiral enhancement required to simultaneously explain the $(g-2)_{e}$ and $(g-2)_{\mu}$ data with either a single scalar leptoquark or a pair of scalar leptoquarks. We provide classification of these scenarios in terms of their ability to satisfy the existing limits on the branching ratio for the $\mu \to e \gamma$ process. The simultaneous explanation of the $(g-2)_{e,\mu}$ discrepancies, coupled with the current experimental data, implies that the $(g-2)_e$ loops are exclusively due to the charm quark propagation whereas the $(g-2)_\mu$ loops are due to the top quark propagation. The scenarios where the $(g-2)_e$ loops are due to the top (bottom) quark propagation are, at best, approximately nine (three) orders of magnitude away from the experimental limit on the $\mu \to e \gamma$ branching ratio. All in all, there are only three particular scenarios that can pass the $\mu \to e \gamma$ test and simultaneously create large enough impact on the $(g-2)_{e,\mu}$ discrepancies when the new physics is based on the Standard Model fermion content. These are the $S_1$, $R_2$, and $S_1 \& S_3$ scenarios, where the first two are already known to be phenomenologically viable candidates with respect to all other flavor and collider data constraints. We show that the third scenario, where the right-chiral couplings to charged leptons are due to $S_1$, the left-chiral couplings to charged leptons are due to $S_3$, and the two leptoquarks mix through the Standard Model Higgs field, cannot address the $(g-2)_{e}$ and $(g-2)_{\mu}$ discrepancies at the $1\sigma$ level due to an interplay between $K_L^0 \to e^\pm \mu^\mp$, $Z \to e^+ e^-$, and $Z \to \mu^+ \mu^-$ data despite the ability of that scenario to avoid the $\mu \to e \gamma$ limit.
\end{abstract}

\newpage
\tableofcontents
\setcounter{footnote}{0}

\section{Introduction}
The observed anomalous magnetic moments of electrons and muons exhibit tension with the corresponding Standard Model (SM) predictions. In particular, the experimental results ($a_{e,\mu}^\mathrm{exp}$) for the electron~\cite{Parker:2018vye} and muon~\cite{Bennett:2006fi} anomalous magnetic moments deviate from the SM predictions ($a_{e,\mu}^\mathrm{SM}$) roughly at the $3$\,$\sigma$~\cite{Aoyama:2017uqe} and $4$\,$\sigma$~\cite{Jegerlehner:2009ry, Keshavarzi:2018mgv, Jegerlehner:2018gjd, Davier:2019can, Aoyama:2020ynm} levels, respectively. More precisely, the observed discrepancies that are of opposite signs currently read
\begin{align}
\label{eq:Delta_a}
&\Delta a_{e}= a_{e}^\mathrm{exp} - a_{e}^\mathrm{SM} =-(8.7\pm 3.6)\times 10^{-13},\\
\label{eq:Delta_b}
&\Delta a_{\mu}= a_\mu^\mathrm{exp} - a_\mu^\mathrm{SM} = (2.79\pm 0.76)\times 10^{-9}.
\end{align}

Various sources of new physics are known to be capable to substantially alter the SM values for $a_{e}=(g-2)_{e}/2$ and $a_{\mu}=(g-2)_{\mu}/2$. (For a sample of studies that analyse effects of the new physics sources on anomalous magnetic moments see, for example, Refs.~\cite{Giudice:2012ms, Queiroz:2014zfa, Biggio:2014ela,ColuccioLeskow:2016dox, Crivellin:2017dsk, Davoudiasl:2018fbb,Crivellin:2018qmi,Liu:2018xkx,Dutta:2018fge, Han:2018znu, Crivellin:2019mvj, Endo:2019bcj, Abdullah:2019ofw, Bauer:2019gfk, Badziak:2019gaf, Mandal:2019gff, Hiller:2019mou, CarcamoHernandez:2019ydc, Cornella:2019uxs, Endo:2020mev, CarcamoHernandez:2020pxw, Haba:2020gkr, Jana:2020pxx,Calibbi:2020emz, Chen:2020jvl, Yang:2020bmh, Hati:2020fzp, Dutta:2020scq, Botella:2020xzf, Chen:2020tfr}.) We, in this study, classify all potentially viable scenarios to explain both discrepancies with either one or two scalar leptoquarks using, as the main tool, the level of disagreement with the existing limits on the branching ratio for the $\mu \to e \gamma$ process. Our analysis dovetails the approach of Ref.~\cite{Lindner:2016bgg} and builds upon the results presented in Refs.~\cite{Crivellin:2018qmi, Bigaran:2020jil}. 

There are only four scalar leptoquark multiplets one needs to consider, as we demonstrate later on, if the new physics scenarios for $a_{e}$ and $a_{\mu}$ are based on the SM fermionic content and include up to two scalar leptoquarks. These leptoquarks are $S_3(\overline{\mathbf{3}},\mathbf{3},1/3)$, $R_2(\mathbf{3},\mathbf{2},7/6)$, $\widetilde{R}_2(\mathbf{3},\mathbf{2},1/6)$, and $S_1(\overline{\mathbf{3}},\mathbf{1},1/3)$, where we specify the transformation properties of  leptoquarks under the SM gauge group $SU(3) \times SU(2) \times U(1)$. (For the reviews on the leptoquark physics see, for example, Refs.~\cite{Buchmuller:1986zs,Dorsner:2016wpm}.) In fact, it has been recently shown in Ref.~\cite{Bigaran:2020jil} that it is possible to simultaneously explain both discrepancies with the new physics that is generated by a single leptoquark extension of the SM, where the leptoquark in question is either $S_1$ or $R_2$. In view of these promising results our study aims to clarify whether there are any additional scenarios that might accomplish the same objective.

To address discrepancies between predicted and observed values for $(g-2)_{e}$ and $(g-2)_{\mu}$ with scalar leptoquarks it is necessary that the associated one-loop level contributions are the quark mass chirality enhanced to be of sufficient strength. The only leptoquarks that can couple to leptons of both chiralities, which is a prerequisite for such an enhancement, are $S_1$ and $R_2$ and the SM gauge symmetry dictates the presence of the up-type quarks in the loop~\cite{Cheung:2001ip}. The only significant difference between the $S_1$ and $R_2$ mediations is that the former is due to the $Q=1/3$ leptoquark while the latter is due to the $Q=5/3$ one, where $Q$ denotes electric charge in units of absolute value of the electron charge. It has been established that one needs the top quarks in the $(g-2)_{\mu}$ loops if one is to address the observed discrepancy at the $1$\,$\sigma$ level and still be in agreement with the ever more stringent combination of constraints from the flavor physics experiments and LHC~\cite{Kowalska:2018ulj}. The possibility to address the $(g-2)_{\mu}$ discrepancy with the top quark chirality enhanced loop contribution is phenomenologically viable even if one resorts to a scenario when two leptoquarks of the same electric charge couple to the muon-top quark pairs of opposite chiralities while mixing with each other through the Higgs field in order to close the loop~\cite{Dorsner:2019itg}. This corresponds to the scenario when $S_1$ mixes with the $Q=1/3$ component of $S_3$, where $S_1$ provides the right-chiral couplings to leptons while $S_3$ provides the left-chiral ones. The chirality enhanced loops for $(g-2)_{e}$ that are generated by $S_1$, $R_2$, or the $S_1 \& S_3$ combination, on the other hand, can be closed not only with the top quarks but also with the charm quarks without any conflict with the existing experimental limits such as those due to the $D$-$\bar{D}$ oscillation and/or atomic parity violation measurements for the leptoquark masses that are allowed by the LHC data analyses. Moreover, in the two leptoquark scenario based on the $R_2 \& \widetilde{R}_2$ combination it is possible to close the $(g-2)_{e}$ loops in a phenomenologically viable way with the bottom quarks~\cite{Dorsner:2019itg} as well.

We systematically study the ability of new physics scenarios with up to two scalar leptoquarks to simultaneously accommodate the $(g-2)_{e,\mu}$ discrepancies using, as the primary classification tool, the experimental limit originating from the $\mu \to e \gamma$ process. We find that there is only one additional scenario, beside the $S_1$ and $R_2$ scenarios that have already been discussed in Ref.~\cite{Bigaran:2020jil}, that is not constrained by this limit. The scenario in question corresponds to the $S_1 \& S_3$ combination, where the left-chiral couplings are provided solely by $S_3$ and the two leptoquarks mix through the SM Higgs field. We furthermore show that the $S_1 \& S_3$ scenario cannot address the $(g-2)_{e,\mu}$ discrepancies at the $1\,\sigma$ level due to an interplay between the $K_L^0 \to e^\pm \mu^\mp$, $Z \to e^+ e^-$, and $Z \to \mu^+ \mu^-$ data even though it clears the $\mu \to e \gamma$ hurdle. In all instances we always turn on four Yukawa couplings between the leptoquark(s) and the relevant quark-lepton pairs to be able to generate chirality enhanced contributions of sufficient strength. We summarise in Table~\ref{tab:LQ_scenarios} the scenarios we consider to simultaneously address the $(g-2)_{e,_\mu}$ discrepancies as well as the associated predictions for the lowest possible values of $Br\left(\mu\to e\gamma\right)$.
\begin{table}[h]
\centering
\begin{tabular}{|c|c|c|c|}
\hline
SCENARIO &  $(g-2)_e$ & $(g-2)_\mu$ & $(Br\left(\mu\to e\gamma\right))^\mathrm{min}$\\
\hline
\hline 
$S_1:\,(q,\,Q)$ & $(t,\,1/3)$ & $(t,\,1/3)$ & $\frac{\tau_\mu \alpha m_\mu^3}{8}
\frac{|\Delta a_e \Delta a_\mu|}{m_e m_\mu}
$\\
$S_1:\,(q,\,Q)$ & $(c,\,1/3)$ & $(t,\,1/3)$ & $0$\\
$R_2:\,(q,\,Q)$ & $(t,\,5/3)$ & $(t,\,5/3)$ & $\frac{\tau_\mu \alpha\; m_\mu^3}{8}
\frac{|\Delta a_e \Delta a_\mu|}{m_e m_\mu}
$\\
$R_2:\,(q,\,Q)$ & $(c,\,5/3)$ & $(t,\,5/3)$ & $0
$\\
$S_1 \& S_3:\,(q,\,Q)$ & $(t,\,1/3)$ & $(t,\,1/3)$ & $\frac{\tau_\mu \alpha\; m_\mu^3}{8}
\frac{|\Delta a_e \Delta a_\mu|}{m_e m_\mu}
$\\
$S_1 \& S_3:\,(q,\,Q)$ & $(c,\,1/3)$ & $(t,\,1/3)$ & $0
$\\
$R_2 \& \widetilde{R}_2:\,(q,\,Q)$ & $(b,\,2/3)$ & $(t,\,5/3)$ & $\frac{\tau_\mu \alpha\; m_\mu^3 \pi^2}{m_e^2 m_\mu^2 m_t^2} \frac{|V_{tb}|^2}{|V_{cb}|^2}
\frac{|\Delta a_e \Delta a_\mu|^2}{(1+4 \ln x_t)^2} M^4
$\\
 \hline
\end{tabular}
\caption{\label{tab:LQ_scenarios} The leptoquark scenarios that have the potential to simultaneously address the $(g-2)_{e,_\mu}$ discrepancies and the associated minimal value of the $\mu\to e\gamma$ branching ratio $(Br\left(\mu\to e\gamma\right))^\mathrm{min}$. We specify for each scenario the quark $q$ that is behind the chirality enhanced contribution and the electric charge $Q$ of the leptoquark in the loops. See the text for details on the notation.}
\end{table}

The paper is organized as follows. In Sec.~\ref{sec:g-2-lq} we elaborate on the scalar leptoquark contributions towards anomalous magnetic moments of electrons and muons and discuss the associated effect on the $\mu \to e \gamma$ process that we find to be the origin of the most relevant flavor constraint. We then proceed to discuss abilities of four different scenarios to simultaneously explain the $(g-2)_{e}$ and $(g-2)_{\mu}$ discrepancies with scalar leptoquarks. The first two scenarios that rely on the single leptoquark contributions towards both anomalous magnetic moments in question are discussed in Sec.~\ref{sec:g-2-single} while the remaining two possibilities are addressed in Sec.~\ref{sec:g-2-pair}. We summarize our findings in Sec.~\ref{sec:conclusion}.

\section{Addressing \texorpdfstring{$(g-2)_{e,\mu}$}{(g-2)e mu} with scalar leptoquarks}
\label{sec:g-2-lq}

We first present an overview of the scalar leptoquark effects on $(g-2)_{e,\mu}$ and $\mu \to e \gamma$ process using the $S_1$ scenario for concreteness. The Yukawa couplings of $S_1$ are~\cite{Buchmuller:1986zs}
\begin{align}
\label{LagS1_a}
\mathcal{L} \supset y_{ij}^L\; \overline{Q^c}_i i\sigma_2 S_1 L_j +y_{ij}^R\; \overline{u^c_R}_i  S_1 \ell_{R j} 
+ \text{h.c.},
\end{align}
where $Q_i =(u_{L i} \quad d_{L i})^T$ and $L_j =(\nu_{L i} \quad \ell_{L i})^T$ are the left-chiral quark and lepton $SU(2)$ doublets, $u_{R i}$, and $\ell_{R j}$ are the right-chiral up-type quarks and charged leptons, respectively, $\sigma_2$ is the Pauli matrix, and $i,j = 1,2,3$ are flavor indices. The Yukawa coupling matrices $y^L$ and $y^R$ are a priori arbitrary $3\times 3$ matrices in the flavor space. The $S_{1}$ diquark couplings have been omitted to ensure proton stability. 

To calculate the flavor observables, it is convenient to rewrite the Lagrangian of Eq.~\eqref{LagS1_a} in the SM fermion mass eigenbasis, to which end we implement the following unitary transformations of the SM fermion fields: $u_L \to U_L u_L$, $u_R \to U_R u_R$, $d_L \to D_L d_L$, $d_R \to D_R d_R$, $\ell_L \to E_L \ell_L$, $\ell_R \to E_R \ell_R$, and $\nu_L \to N_L \nu_L$. These transformations represent the most general redefinitions of both the left-chiral and the right-chiral fields as long as one defines $U= E_L^\dagger N_L$ and $V= U_L^\dagger D_L$, where $U$ and $V$ are the Pontecorvo-Maki-Nakagawa-Sakata and Cabibbo-Kobayashi-Maskawa matrices, respectively. Note that the unitary transformations of the right-chiral fermions are not physical in the SM whereas that might not be the case if one considers its extensions. With these redefinitions, the part of Lagrangian presented in Eq.~\eqref{LagS1_a} takes the following form:
\begin{align}
\label{LagS1}
&\mathcal{L}_{S_1} = -(D_L^T y^L N_L)_{ij} \overline{d^c_L}_i S_1 \nu_{Lj} + (U_L^T y^L E_L)_{i j} \overline{u^c_L}_i S_1 \ell_{Lj} + (U_R^T y^R E_R)_{ij}\overline{u^c_R}_i S_1 \ell_{R j} + \text{h.c.}.  
\end{align}

There are two particular bases that are commonly used in the literature to study flavor physics signatures. One would correspond to the case when the CKM rotations are entirely in the down-type quark sector. We will refer to it as the up-type quark mass-diagonal basis. The other basis corresponds to the case when the CKM rotations are purely in the up-type quark sector and we accordingly refer to it as the down-type quark mass-diagonal basis. We will, in what follows, always specify both the Yukawa ansatz and the unitary transformations that we are going to use to study a given new physics scenario. The former will be given in the flavor basis while the latter will be specified in the mass eigenstate basis. In practical terms this amounts to specifying Yukawa couplings using Eq.~\eqref{LagS1_a} and the unitary transformations using Eq.~\eqref{LagS1}. Note that in our approach the predictions will change if we keep the same Yukawa ansatz but work with different unitary transformations. 

Since the right-chiral rotations will not be relevant for our analysis it will be understood that $U_R=D_R=E_R=I$ throughout the rest of this manuscript, where $I$ is the identity matrix. Moreover, we will always take the PMNS rotations to be in the neutrino sector. The up-type quark mass-diagonal basis is thus implemented via $U_L=U_R=D_R=E_L=E_R=I$, whereas the down-type quark mass-diagonal basis is specified through $U_R=D_L=D_R=E_L=E_R=I$. Consequently, the up-type quark mass-diagonal basis corresponds to the $D_L \equiv V$ case while the down-type quark mass-diagonal basis is give through $U_L \equiv V^\dagger$. Of course, there is a continuous set of unitary transformations that would take one from the first basis into the second basis and vice versa. And, in our approach, each of these transformations would yield, for the same Yukawa coupling ansatz, its own phenomenological signatures.

It is now possible, using lagrangian of Eq.~\eqref{LagS1}, to write down the $S_1$ contributions towards $(g-2)_{e,\mu}$. We will accomplish this by switching on, in Eq.~\eqref{LagS1_a}, Yukawa couplings $y^L_{32}$, $y^L_{31}$, $y^R_{32}$, and $y^R_{31}$, and working in the down-type quark mass-diagonal basis. This scenario corresponds to the situation when $S_1$ couples simultaneously to the electron-top quark and muon-top quark pairs. We accordingly obtain 
\begin{align}
\label{eq:Deltaa_e}
&\Delta a_e=-\frac{3 m^2_e}{8\pi^2 M^2}
\left[ 
\frac{m_t}{m_e} Re\left( V^*_{tb}y^L_{31}(y^R_{31})^* \right)
\left( \frac{7}{6}+\frac{2}{3}\ln x_t \right) - \frac{1}{12} \left( |y^R_{31}|^2+|y^L_{31}|^2   \right) 
\right],
\\
\label{eq:Deltaa_mu}
&\Delta a_\mu=-\frac{3 m^2_\mu}{8\pi^2 M^2}
\left[ 
\frac{m_t}{m_\mu} Re\left( V^*_{tb}y^L_{32}(y^R_{32})^* \right)
\left( \frac{7}{6}+\frac{2}{3}\ln x_t \right) - \frac{1}{12} \left( |y^R_{32}|^2+|y^L_{32}|^2   \right) 
\right].
\end{align}
Clearly, it is necessary to switch on at least four Yukawa couplings to simultaneously affect $(g-2)_{e}$ and $(g-2)_{\mu}$ with the chirality enhanced contributions. One pair of couplings enters $(g-2)_{e}$ and the other $(g-2)_{\mu}$. Note that we define, for convenience, $x_t=m^2_t/M^2$, where $m_t$ is the top quark mass and $M$ is the mass of $S_1$ leptoquark.

The current limit on the branching ratio for $\mu\to e \gamma$ process is $Br\left(\mu\to e\gamma\right)<4.2\times 10^{-13}$~\cite{TheMEG:2016wtm}. We find it to be the most severe obstacle to simultaneous explanation of the $(g-2)_{e,\mu}$ discrepancies with the scalar leptoquark physics. For example, if  the leading parts of the $(g-2)_{e,\mu}$ loops are proportional to the top quark mass, as given in Eqs.~\eqref{eq:Deltaa_e} and \eqref{eq:Deltaa_mu}, the new physics contribution towards $\mu\to e \gamma$ is \cite{Lavoura:2003xp}
\begin{align}
&Br(\mu\to e \gamma)=\frac{9\alpha \tau_\mu m^5_\mu}{1024\pi^4M^4}
\left( |A_1|^2 +|B_1|^2 \right),
\end{align}
where
\begin{align}
&A_1=-\frac{1}{12}\left[ y^L_{32}(y^L_{31})^* +\frac{m_e}{m_\mu} y^R_{32}(y^R_{31})^*  \right]
+\frac{m_t}{m_\mu}\left( \frac{7}{6}+\frac{2}{3}\ln x_t \right) V_{tb} y^R_{32}(y^L_{31})^*,\label{A}
\\
&B_1=-\frac{1}{12}\left[ y^R_{32}(y^R_{31})^* +\frac{m_e}{m_\mu} y^L_{32}(y^L_{31})^*  \right]
+\frac{m_t}{m_\mu}\left( \frac{7}{6}+\frac{2}{3}\ln x_t \right) V_{tb}^* y^L_{32}(y^R_{31})^*.\label{B}
\end{align}

There are also new physics contributions to other processes such as $\mu$-$e$ conversion, $\mu \to eee$, and $Z\to \ell \ell^{'}$, to name a few, that are generated once one tries to simultaneously address $(g-2)_{e}$ and $(g-2)_{\mu}$ with scalar leptoquarks. We take into account these constraints only if they are not subdominant with respect to the $\mu \to e \gamma$ limit in what follows. 

\subsection{Single leptoquark scenarios: \texorpdfstring{$S_1$}{S1} and \texorpdfstring{$R_2$}{R2}} 
\label{sec:g-2-single}

\subsubsection{\texorpdfstring{$S_1$}{S1} with the top quark loops}

Let us start with the $S_1$ case, when the $(g-2)_{e}$ and $(g-2)_{\mu}$ loops are both top quark induced, and with real Yukawa couplings $y^L_{31}$, $y^L_{32}$, $y^R_{31}$, and $y^R_{32}$, as defined in Eq.~\eqref{LagS1_a}, switched on. We will work in the down-type quark mass-diagonal basis in which the CKM resides in the up-type quark sector. Again, the PMNS rotations are taken to reside in the neutrino sector while the unitary rotations of the right-chiral fermions are assumed to be identity matrices throughout this manuscript.

The leading chirality enhanced contributions towards $\Delta a_e$ and $\Delta a_\mu$ are 
\begin{align}
\label{eq:S1_a}
&\Delta a_e=-\frac{3}{8\pi^2}\frac{m_tm_e}{M^2}y^L_{31}y^R_{31} \left( \frac{7}{6}+\frac{2}{3}\ln x_t \right), 
\\
\label{eq:S1_b}
&\Delta a_\mu=-\frac{3}{8\pi^2}\frac{m_tm_\mu}{M^2}y^L_{32}y^R_{32} \left( \frac{7}{6}+\frac{2}{3}\ln x_t \right),  
\end{align} 
while the $\mu\to e \gamma$ contribution is
\begin{equation}
\label{eq:S1_c}
Br(\mu\to e\gamma)=\frac{\tau_\mu \alpha\; m_\mu^3}{4} \left| 
\frac{3m_t}{16\pi^2 M^2} \left( \frac{7}{6}+\frac{2}{3}\ln x_t \right)
\right|^2
\left[  \left|y^R_{31}y^L_{32}\right|^2 + \left|y^L_{31}y^R_{32}\right|^2 \right].
\end{equation}
If we define $x=y^R_{31}/y^R_{32}$ and rearrange Eqs.~\eqref{eq:S1_a}, \eqref{eq:S1_b}, and \eqref{eq:S1_c}, we obtain the following expression for $Br(\mu\to e\gamma)$ in terms of $\Delta a_e$ and $\Delta a_\mu$:
\begin{equation}
\label{eq:no_go_S1_t_t}
Br(\mu\to e\gamma)=
\frac{\tau_\mu \alpha\; m_\mu^3}{16}
\left(
\frac{\Delta a^2_e}{m^2_e}\;\frac{1}{x^2}+
\frac{\Delta a^2_\mu}{m^2_\mu}\;x^2
\right).
\end{equation}
An especially nice feature of the prediction for $Br(\mu\to e\gamma)$, as given in Eq.~\eqref{eq:no_go_S1_t_t}, is that it does not exhibit dependence on the scale of new physics whatsoever. If we determine the minimal value of the right-hand side of Eq.~\eqref{eq:no_go_S1_t_t} with respect to $x^2$ that we denote with $(Br(\mu\to e\gamma))^\mathrm{min}$ we obtain
\begin{equation}
\label{eq:no_go_S1_t_t_b}
(Br(\mu\to e\gamma))^\mathrm{min}=\frac{\tau_\mu \alpha\; m_\mu^3}{8}
\frac{|\Delta a_e \Delta a_\mu|}{m_e m_\mu}= 1.6 \times 10^{-4},
\end{equation}
where the central values for $\Delta a_e$ and $\Delta a_\mu$, as given in Eqs.~\eqref{eq:Delta_a} and~\eqref{eq:Delta_b}, respectively, are inserted for convenience. We also use $\tau_\mu=3.33941 \times 10^{18}$\,GeV$^{-1}$, $\alpha=1/137$, $m_\mu= 105.65$\,MeV, and $m_e= 0.5109$\,MeV~\cite{Tanabashi:2018oca}. 

The prediction for the minimal attainable value $(Br(\mu\to e\gamma))^\mathrm{min}$ in Eq.~\eqref{eq:no_go_S1_t_t_b} has been obtained in Ref.~\cite{Crivellin:2018qmi} via the effective field theory approach, under the assumption that the single source of new physics couples simultaneously to the muon-top quark and electron-top quark pairs. The $(Br(\mu\to e\gamma))^\mathrm{min}$ value, obtained for $(x^2)^\mathrm{min}=(|\Delta a_e | m_\mu)/(|\Delta a_\mu| m_e)$, makes it transparent that it is impossible to reconcile the experimental limit on $Br(\mu\to e\gamma)$ with required shifts in $\Delta a_e$ and $\Delta a_\mu$ for any value of $x$ since the predicted minimal value and the experimentally observed limit are already, at best, eight to nine orders of magnitude apart. The same conclusion holds if one evaluates these observables for the same Yukawa ansatz, i.e., $y^L_{31}, y^L_{32}, y^R_{31}, y^R_{32} \neq 0$, but in the up-type quark mass-diagonal basis. We accordingly quote this result for $(Br(\mu\to e\gamma))^\mathrm{min}$ in Table~\ref{tab:LQ_scenarios}.

\subsubsection{\texorpdfstring{$S_1$}{S1} with the top and charm quark loops} 

One might entertain a possibility to address $(g-2)_{e}$ with the charm quark loops and $(g-2)_{\mu}$ with the top quark loops. If we work in the down-type quark mass-diagonal basis and switch on real Yukawa couplings $y^L_{21}$, $y^L_{32}$, $y^R_{21}$, and $y^R_{32}$, as defined in Eq.~\eqref{LagS1_a}, we obtain the following expressions for $\Delta a_e$, $\Delta a_\mu$, and $Br(\mu\to e\gamma)$:
\begin{align}
\label{eq:S1_aa}
&\Delta a_e= - \frac{3m_e m_c}{8\pi^2 M^2} \left( \frac{7}{6}+\frac{2}{3}\ln x_c \right)V_{cs} y^L_{21}y^R_{21},
\\
\label{eq:S1_bb}
&\Delta a_\mu = - \frac{3m_\mu m_t}{8\pi^2 M^2} \left( \frac{7}{6}+\frac{2}{3}\ln x_t \right)V_{tb} y^L_{32}y^R_{32},
\\
\label{eq:S1_cc}
&Br(\mu\to e\gamma)=\frac{9\alpha \tau_\mu m^5_\mu}{1024\pi^4 M^4}
\left[ |V_{ts}|^2
\frac{m^2_t}{m^2_\mu}\left( \frac{7}{6}+\frac{2}{3}\ln x_t \right)^2 \left( y^R_{32}y^L_{21} \right)^2
\right.  \nonumber \\& \left. \hspace{5cm}
+ |V_{cb}|^2\frac{m^2_c}{m^2_\mu}\left( \frac{7}{6}+\frac{2}{3}\ln x_c \right)^2 \left( y^L_{32}y^R_{21} \right)^2
\right],
\end{align}
where we introduce $x_c=m^2_c/M^2$ with $m_c$ being the charm quark mass and neglect the subleading contributions in $\Delta a_e$ and $\Delta a_\mu$. Note that it is the CKM matrix that induces the $\mu\to e\gamma$ process within this basis and with this particular Yukawa ansatz.

If we rearrange Eqs.~\eqref{eq:S1_aa}, \eqref{eq:S1_bb}, and \eqref{eq:S1_cc}, and define $x=y^R_{21}/y^R_{32}$, we obtain 
\begin{equation}
\label{eq:no_go_S1_t_c}
Br(\mu\to e\gamma)=
\frac{\tau_\mu \alpha\; m_\mu^3}{16}
\left(
\frac{\Delta a^2_e}{m^2_e}\;\frac{\widetilde{A}}{x^2} \frac{|V_{ts}|^2}{|V_{cs}|^2}+
\frac{\Delta a^2_\mu}{m^2_\mu}\;\frac{x^2}{\widetilde{A}} \frac{|V_{cb}|^2}{|V_{tb}|^2}
\right),
\end{equation}
where
\begin{equation}
\widetilde{A}= \frac{m^2_t}{m^2_c} \frac{\left( \frac{7}{6}+\frac{2}{3}\ln x_t \right)^2}{\left( \frac{7}{6}+\frac{2}{3}\ln x_c \right)^2}.
\end{equation}

This time around the expression for $Br(\mu\to e\gamma)$, as given in Eq.~\eqref{eq:no_go_S1_t_c}, exhibits logarithmic dependence on the new physics scale but the minimal attainable value does not. Namely, we obtain that 
\begin{equation}
\label{eq:no_go_S1_t_c_b}
(Br(\mu\to e\gamma))^\mathrm{min}=
\frac{\tau_\mu \alpha\; m_\mu^3}{8}
\frac{|\Delta a_e \Delta a_\mu|}{m_e m_\mu}
\frac{|V_{ts}||V_{cb}|}{|V_{cs}||V_{tb}|}= 2.8 \times 10^{-7},
\end{equation}
where we use $|V_{cs}|=0.9735$, $|V_{tb}|=0.9991$, $|V_{cb}|=0.0416$, and $|V_{ts}|=0.0409$~\cite{Tanabashi:2018oca}. We derive Eq.~\eqref{eq:no_go_S1_t_c_b} from Eq.~\eqref{eq:no_go_S1_t_c} for $(x^2)^\mathrm{min}=\widetilde{A} (|\Delta a_e | m_\mu |V_{ts}||V_{tb}|)/(|\Delta a_\mu| m_e |V_{cs}||V_{cb}|)$. 

It is clear from Eq.~\eqref{eq:no_go_S1_t_c_b} that this particular $S_1$ scenario fails to reconcile required shifts in $\Delta a_e$ and $\Delta a_\mu$ with the current bound on $Br(\mu\to e\gamma)$. The minimal predicted value for the branching ratio for $\mu\to e\gamma$ is the CKM matrix suppressed with respect to the case when both anomalous magnetic moment discrepancies are addressed with the top quark loops. Nevertheless, the minimal predicted value is still six orders of magnitude away from the associated $Br(\mu\to e\gamma)$ experimental limit.

It has been noted recently in Ref.~\cite{Bigaran:2020jil} that if one takes the same Yukawa ansatz, i.e., $y^L_{21}, y^L_{32}, y^R_{21}, y^R_{32} \neq 0$, but works instead in the up-type quark mass-diagonal basis, one completely suppresses the $\mu\to e\gamma$ signature since $S_1$ couples separately to the muon-top quark and electron-charm quark pairs. One can recreate the up-type quark mass-diagonal basis results for $\Delta a_e$, $\Delta a_\mu$, and $Br(\mu\to e\gamma)$ by setting $|V_{ts}|=|V_{cb}|=0$ and $|V_{cs}|=|V_{tb}|=1$ in Eqs.~\eqref{eq:S1_aa}, \eqref{eq:S1_bb}, and \eqref{eq:S1_cc}, respectively. The authors of Ref.~\cite{Bigaran:2020jil} have subsequently demonstrated that the $S_1$ scenario that addresses $(g-2)_{e}$ with the charm quark loops and $(g-2)_{\mu}$ with the top quark loops is phenomenologically viable with respect to all current experimental data. We accordingly quote that $(Br(\mu\to e\gamma))^\mathrm{min}=0$ for this particular scenario in Table~\ref{tab:LQ_scenarios}.

\subsubsection{\texorpdfstring{$R_2$}{R2} with the top quark loops}

The analysis of the $R_2$ scenario, where the $(g-2)_{e,\mu}$ loops are both generated with the top quark, will mirror the $S_1$ case, as we show next. The relevant part of the Lagrangian is
\begin{align}
\label{eq:R2}
\mathcal{L} \supset -y_{ij}^L \overline{u}_{R i} R_2 i\sigma_2 L_{L_{j}} + y_{ij}^R \overline{Q}_{L i} R_2\ell_{Rj}\,  + \text{h.c.},
\end{align}
where $y^L$ and $y^R$ are the Yukawa coupling matrices associated with $R_2$. If we go to the mass and electric charge eigenstate basis we have that
\begin{eqnarray}
\nonumber
\mathcal{L}_{R_2} &=& - (U^\dagger_R y^L E_L)_{i j} \overline{u}_{R i} \ell_{L j} R_2^{5/3} + (U_L^\dagger y^R E_R)_{i j} \, \overline{u}_{Li} \ell_{Rj} R_2^{5/3}\\
 && + (U^\dagger_R y^L N_L)_{i j} \overline{u}_{Ri} \nu_{Lj} R_2^{2/3}  + (D_L^\dagger y^R E_R)_{i j} \overline{d}_{Li} \ell_{Rj} R_2^{2/3} + \text{h.c.}, 
\end{eqnarray}
where $R_2^{5/3}$ and $R_2^{2/3}$ are $Q=5/3$ and $Q=2/3$ components of $R_2$ multiplet, respectively. We will assume that both components of $R_2$ are mass degenerate and denote the corresponding mass with $M$ in what follows.

To generate $\Delta a_\ell$ contributions with the chirality enhanced top quark loops we need to switch on $y^R_{3\ell}$ and $y^L_{3\ell}$, as defined in Eq.~\eqref{eq:R2}, where $\ell=1,2=e,\mu$. In the down-type quark mass-diagonal basis this Yukawa ansatz yields  
\begin{align}
&\Delta a_\ell=-\frac{3 m^2_\ell}{8\pi^2 M^2}
\left[ 
\frac{m_t}{m_\ell} Re\left[ (V_{tb}y^R_{3\ell})^* y^L_{3\ell} \right]
\left( \frac{1}{6}+\frac{2}{3}\ln x_t \right) + \frac{1}{4} \left( |y^R_{3\ell}|^2+|y^L_{3\ell}|^2   \right) 
\right],\\
\label{eq:R2_c}
&Br(\mu\to e \gamma)=\frac{9\alpha \tau_\mu m^5_\mu}{1024\pi^4M^4}
\left( |A_2|^2 +|B_2|^2 \right),
\end{align}
where
\begin{align}
&A_2=\frac{1}{4}\left[ y^R_{32}(y^R_{31})^* +\frac{m_e}{m_\mu} y^L_{32}(y^L_{31})^*  \right]
+\frac{m_t}{m_\mu}\left( \frac{1}{6}+\frac{2}{3}\ln x_t \right) (V_{tb} y^R_{31})^* y^L_{32},
\\
&B_2=\frac{1}{4}\left[ y^L_{32}(y^L_{31})^* +\frac{m_e}{m_\mu} y^R_{32}(y^R_{31})^*  \right]
+\frac{m_t}{m_\mu}\left( \frac{1}{6}+\frac{2}{3}\ln x_t \right) (y^L_{31})^* V_{tb} y^L_{32}.
\end{align}
The expression for $Br(\mu\to e \gamma)$ in Eq.~\eqref{eq:R2_c} translates into the exact same functional form, including the numerical prefactors, as that of Eq.~\eqref{eq:no_go_S1_t_t}, where, again, we define $x=y^R_{31}/y^R_{32}$ but, this time, for the $R_2$ Yukawa couplings. That accordingly yields the same minimal value for $Br(\mu\to e \gamma)$ as given in Eq.~\eqref{eq:no_go_S1_t_t_b} and thus shows that the $R_2$ scenario with the chirality enhanced top quarks loops when the CKM matrix resides in the up-type quark sector also fails to simultaneously address the $(g-2)_{e,\mu}$ discrepancies in phenomenologically viable way due to the conflict with the $\mu\to e \gamma$ constraint. The same conclusion holds if one works in the up-type quark mass-diagonal basis and agrees with the the effective field theory approach results of Ref.~\cite{Crivellin:2018qmi}. The value we quote in Table~\ref{tab:LQ_scenarios} for $(Br(\mu\to e\gamma))^\mathrm{min}$ reflects that fact. 

\subsubsection{\texorpdfstring{$R_2$}{R2} with the top and charm quark loops} 

If we try to accommodate $(g-2)_{e}$ with the charm quark loops and $(g-2)_{\mu}$ with the top quark loops by switching on, in Eq.~\eqref{eq:R2}, $y^R_{21}$, $y^L_{21}$, $y^R_{32}$, and $y^L_{32}$, while working in the down-type quark mass-diagonal basis, we find that
\begin{align}
\label{eq:R2_aa}
&\Delta a_e= - \frac{3m_e m_c}{8\pi^2 M^2} \left( \frac{1}{6}+\frac{2}{3}\ln x_c \right)V_{cs} y^R_{21}y^L_{21},
\\
\label{eq:R2_bb}
&\Delta a_\mu = - \frac{3m_\mu m_t}{8\pi^2 M^2} \left( \frac{1}{6}+\frac{2}{3}\ln x_t \right)V_{tb} y^R_{32}y^L_{32},
\end{align}
while
\begin{align}
\label{eq:R2_cc}
&Br(\mu\to e\gamma)=\frac{9\alpha \tau_\mu m^5_\mu}{1024\pi^4 M^4}
\left[ |V_{ts}|^2
\frac{m^2_t}{m^2_\mu}\left( \frac{1}{6}+\frac{2}{3}\ln x_t \right)^2 \left( y^R_{21}y^L_{32} \right)^2
\right.  \nonumber \\& \left. \hspace{5cm}
+
|V_{cb}|^2\frac{m^2_c}{m^2_\mu}\left( \frac{1}{6}+\frac{2}{3}\ln x_c \right)^2 \left( y^L_{21}y^R_{32} \right)^2
\right].
\end{align}
If we introduce $x=y^R_{21}/y^R_{32}$ and combine Eqs.~\eqref{eq:R2_aa}, \eqref{eq:R2_bb}, and \eqref{eq:R2_cc}, we obtain the following expression for $Br(\mu\to e\gamma)$ in terms of $\Delta a_e$ and $\Delta a_\mu$: 
\begin{align}
\label{eq:R2tc}
Br(\mu\to e\gamma)&=
\frac{\tau_\mu \alpha\; m_\mu^3}{16}
\left(
\frac{\Delta a^2_e}{m^2_e}\;\frac{1}{x^2} \frac{|V_{cb}|^2}{|V_{cs}|^2}+
\frac{\Delta a^2_\mu}{m^2_\mu}\; x^2 \frac{|V_{ts}|^2}{|V_{tb}|^2}
\right).
\end{align}
It is now trivial to see that Eq.~\eqref{eq:R2tc} yields the same minimal value for the branching ratio for $\mu\to e\gamma$ as given in Eq.~\eqref{eq:no_go_S1_t_c_b}. The fact that we work in the down-type quark mass-diagonal basis within our Yukawa coupling ansatz leads to the CKM matrix induced $\mu\to e\gamma$ process. This thus renders the associated $R_2$ scenario with the chirality enhanced top quarks loops for $(g-2)_{\mu}$ and charm quark loops for $(g-2)_{e}$ not to be phenomenologically viable. If one, on the other hand, keeps the same Yukawa coupling ansatz but adopts the up-type quark mass-diagonal basis the $\mu\to e\gamma$ process gets completely suppressed, as demonstrated in Ref.~\cite{Bigaran:2020jil}. We accordingly quote $(Br(\mu\to e\gamma))^\mathrm{min}=0$ as the minimal value for the $\mu\to e\gamma$ branching ratio for this particular resolution of the $(g-2)_{e,\mu}$ discrepancies in Table~\ref{tab:LQ_scenarios}. The subsequent analysis has shown that it is possible to simultaneously address $(g-2)_e$ with the charm quark loops and $(g-2)_\mu$ with the top quark loops within the $R_2$ scenario without any conflict with the existing experimental data~\cite{Bigaran:2020jil}.

\subsection{Two leptoquark scenarios: \texorpdfstring{$S_1 \& S_3$}{S1 \& S3} and \texorpdfstring{$R_2 \& \widetilde{R}_2$}{R2 \& R2 tilde}} 
\label{sec:g-2-pair}

The two leptoquark scenarios $S_1 \& S_3$ and $R_2 \& \widetilde{R}_2$ open up additional possibilities to simultaneously address the $(g-2)_{e,\mu}$ discrepancies. For example, the $R_2 \& \widetilde{R}_2$ scenario can produce chirality enhanced contributions towards $(g-2)_{e}$ that are proportional to the bottom quark mass. In what follows we systematically go through the various potentially viable possibilities, mirroring the analyses of the single leptoquark scenarios.

\subsubsection{\texorpdfstring{$S_1 \& S_3$}{S1 \& S3} with the top quark loops}
We start with the analysis of the two leptoquark scenario based on the $S_1 \& S_3$ combination. The idea is to address both discrepancies with the chirality enhanced top quark loops, where the leptoquarks in the loop will be mixture of $S_1$ with the $Q=1/3$ state in $S_3$. Our objective is to ascertain the phenomenological viability of this scenario under the assumption that the two leptoquarks couple to charged leptons of opposite chiralities. 

The relevant parts of the new physics Lagrangian that have not been introduced in preceding sections are 
\begin{align}
\label{eq:S3}
\mathcal{L} \supset  y^S_{ij} \overline{Q^c}_{L i} i\sigma_2 (\sigma_a S_3^a) L_{L j} +  \lambda \, H^\dagger (\sigma_a S_3^a)H S_1^\ast+\mathrm{h.c.}\,,
\end{align}
where $\lambda$ is a dimensionless coupling and $\sigma_a$, $a=1,2,3$, are Pauli matrices. It is the second term in Eq.~\eqref{eq:S3} that, after electroweak symmetry breaking, induces a mixing between $S_3^{1/3}$ and $S_1$ via the vacuum expectation value $v$ of the SM Higgs field $H(\mathbf{1},\mathbf{2},1/2)$. Note that the $S_{3}$ diquark couplings have been omitted to ensure proton stability. 

The Yukawa couplings of $S_3$, in the mass eigenstate basis for the SM fermions, are
\begin{eqnarray}
\mathcal{L}_{S_3} &=& -(D_L^T y^S N_L)_{i j} \overline{d^c}_{Li} S_3^{1/3} \nu_{Lj} - \sqrt{2}\; (D_L^T y^S E_L)_{i j} \overline{d^c}_{Li} S_3^{4/3} \ell_{Lj}  \nonumber\\ 
     &&+ \sqrt{2}\; (U_L^T y^S N_L)_{i j} \overline{u^c}_{Li} S_3^{-2/3} \nu_{Lj} - (U_L^T y^S E_L)_{i j} \overline{u^c}_{Li} S_3^{1/3} \ell_{Lj} + \text{h.c.}. 
\label{LYprime}
\end{eqnarray}

We perform the analysis in the leptoquark mass eigenstate basis. The mixing matrix for $S_3^{1/3}$ and $S_1$ is~\cite{Dorsner:2019itg}
\begin{align}
& \begin{pmatrix}
S_-\\S_+
\end{pmatrix}
=\begin{pmatrix}
\cos\theta&\sin\theta\\
-\sin\theta&\cos\theta
\end{pmatrix} \begin{pmatrix}
S_3^{1/3}\\S_1
\end{pmatrix}\equiv \begin{pmatrix}
c_\theta&s_\theta\\
-s_\theta&c_\theta
\end{pmatrix} \begin{pmatrix}
S_3^{1/3}\\S_1
\end{pmatrix},
\end{align}
with
\begin{align}
&\tan2\theta=\frac{\lambda v^2}{M^2_1-M^2_3},
\\
&M^2_{\pm}=\frac{M^2_1+M^2_3}{2}\pm \frac{1}{2} 
\sqrt{ (M^2_1-M^2_3)^2+\lambda^2v^4 },
\end{align}
where $\theta$ is the mixing angle, $M_{\pm}$ are masses of states $S_{\pm}$ with $Q=1/3$, $M_1$ and $M_3$ are masses of $S_1$ and $S_3$ multiplets, respectively, for $\lambda=0$. Note that $S_3^{4/3}$ and $S_3^{-2/3}$ have common mass $M_3$ regardless of the mixing.

We switch on $y^S_{31}$ and $y^S_{32}$, as given in Eq.~\eqref{eq:S3}, as well as $y^R_{31}$, and $y^R_{32}$, as defined in Eq.~\eqref{LagS1_a}, and work in the down-type quark mass-diagonal basis to find that the general formula for $\Delta a_{\ell}$, $\ell=1,2=e,\mu$, in this mixed scenario, reads
\begin{align}
\Delta a_{\ell}=-\frac{3m^2_{\ell}}{8\pi^2}\bigg\{  
&\frac{1}{3M^2_{3}}|y^S_{3\ell}|^2
-\frac{1}{12M^2_-} \left[ |y^R_{3\ell}|^2 s_\theta^2 + |y^S_{3\ell}|^2 c_\theta^2\right]
-\frac{1}{12M^2_+} \left[ |y^R_{3\ell}|^2 c_\theta^2 + |y^S_{3\ell}|^2 s_\theta^2\right]
\nonumber \\
&
+\frac{m_t}{m_{\ell}} Re\left[ (V^*y^S)^*_{3\ell} y^R_{3\ell} \right] 
\left[ \frac{s_{2 \theta}}{2 M^2_+} \left( \frac{7}{6}+\frac{2}{3}\ln x^+_t \right) -\frac{s_{2\theta}}{2 M^2_-}
\left( \frac{7}{6}+\frac{2}{3}\ln x^-_t \right) \right]
\bigg\}.
\end{align}

If we take all the Yukawa couplings to be real and keep only the chirality enhanced terms, we get that
\begin{align}
\label{eq:S1S3_a}
&\Delta a_\ell=-\frac{3m_tm_\ell}{8\pi^2}\;y^R_{3\ell}  y^S_{3\ell}\;\left[ \frac{s_{2\theta}}{2 M^2_+} \left( \frac{7}{6}+\frac{2}{3}\ln x^+_t \right) -\frac{s_{2 \theta}}{2 M^2_-}
\left( \frac{7}{6}+\frac{2}{3}\ln x^-_t \right) \right],
\end{align}
and
\begin{align}
\label{eq:S1S3_c}
Br(\mu\to e\gamma)=\frac{\tau_\mu \alpha\; m_\mu^3}{4}
\frac{9m^2_t}{(16\pi^2)^2}
&\left(
\left|y^R_{32}y^S_{31}\right|^2
+ \left|y^R_{31}y^S_{32}\right|^2
\right)
\nonumber \\&\times 
\left[ \frac{s_{2 \theta}}{2 M^2_+} \left( \frac{7}{6}+\frac{2}{3}\ln x^+_t \right) -\frac{s_{2\theta}}{2 M^2_-}
\left( \frac{7}{6}+\frac{2}{3}\ln x^-_t \right) \right]^2.
\end{align}
If we furthermore define $x=y^R_{31}/y^R_{32}$ and rewrite $Br(\mu\to e\gamma)$ in Eq.~\eqref{eq:S1S3_c} using Eq.~\eqref{eq:S1S3_a} we find that
\begin{equation}
Br(\mu\to e\gamma)=
\frac{\tau_\mu \alpha\; m_\mu^3}{16}
\left(
\frac{\Delta a^2_e}{m^2_e}\;\frac{1}{x^2}+
\frac{\Delta a^2_\mu}{m^2_\mu}\;x^2
\right).
\end{equation}
This expression is identical to the one for the $S_1$ scenario with the top quark loops and accordingly yields the same minimal value for $Br(\mu\to e\gamma)$ as given in Eq.~\eqref{eq:no_go_S1_t_t_b}. Since we obtain the same end result if we work in the up-type quark mass-diagonal basis we conclude that the $S_1 \& S_3$ scenario with the top quark loops is not adequate for the simultaneous explanation of the $\Delta a_e$ and $\Delta a_\mu$ shifts. What we effectively have in both bases is that a single source of new physics simultaneously couples to the muon-top quark and electron-top quark pairs. It is thus not surprising that we obtain the effective field theory approach results of Ref.~\cite{Crivellin:2018qmi}. The value for $(Br(\mu\to e\gamma))^\mathrm{min}$ that we quote in Table~\ref{tab:LQ_scenarios} for this particular $S_1 \& S_3$ scenario reflects that fact. 

\subsubsection{\texorpdfstring{$S_1 \& S_3$}{S1 \& S3} with the top and charm quark loops}

In order to investigate viability of the $S_1 \& S_3$ scenario when the chirality enhanced shift in $(g-2)_\mu$ is generated with the top quark loops and the shift in $(g-2)_e$ is due to the charm quark loops we switch on $y^S_{21}$ and $y^S_{32}$, as given in Eq.~\eqref{eq:S3}, as well as $y^R_{21}$, and $y^R_{32}$, as defined in Eq.~\eqref{LagS1_a}, and work in the down-type quark mass-diagonal basis to find that
\begin{align}
\Delta a_{\ell}=&-\frac{3m^2_{\ell}}{8\pi^2}\bigg\{  
\frac{1}{3M^2_{3}}|y^S_{q\ell}|^2
-\frac{1}{12M^2_-} \left[ |y^R_{q\ell}|^2 s_\theta^2 + |(V^*y^S)_{q\ell}|^2 c_\theta^2\right]
\nonumber \\
&
-\frac{1}{12M^2_+} \left[ |y^R_{q\ell}|^2 c_\theta^2 + |(V^*y^S)_{q\ell}|^2 s_\theta^2\right]
\nonumber \\
&
\label{eq:S1S3tc}
+\frac{m_q}{m_{\ell}} Re\left[ (V^*y^S)^*_{q\ell} y^R_{q\ell}\right] 
\left[ \frac{s_{2\theta}}{2 M^2_+} \left( \frac{7}{6}+\frac{2}{3}\ln x^+_q \right) -\frac{s_{2\theta}}{2 M^2_-}
\left( \frac{7}{6}+\frac{2}{3}\ln x^-_q \right) \right]
\bigg\},
\end{align}
where for index $\ell=e=1$ one needs to set $q=c=2$ while for $\ell=\mu=2$ one needs to take $q=t=3$ when and where appropriate. If Yukawa couplings are real and if we omit subleading terms, Eq.~\eqref{eq:S1S3tc} translates into
\begin{align}
\label{eq:S1S3_t_c_aa}
&\Delta a_e=-\frac{3m_em_c}{8\pi^2}\;V_{cs}y^R_{21}  y^S_{21}\;\left[ \frac{s_{2\theta}}{2 M^2_+} \left( \frac{7}{6}+\frac{2}{3}\ln x^+_c \right) -\frac{s_{2\theta}}{2 M^2_-}
\left( \frac{7}{6}+\frac{2}{3}\ln x^-_c \right) \right],
\\
\label{eq:S1S3_t_c_bb}
&\Delta a_\mu=-\frac{3m_\mu m_t}{8\pi^2}\;V_{tb}y^R_{32}  y^S_{32}\;\left[ \frac{s_{2\theta}}{2 M^2_+} \left( \frac{7}{6}+\frac{2}{3}\ln x^+_t \right) -\frac{s_{2\theta}}{2 M^2_-}
\left( \frac{7}{6}+\frac{2}{3}\ln x^-_t \right) \right].
\end{align}
The branching ratio for $\mu\to e\gamma$ is
\begin{align}
\label{eq:S1S3_t_c_cc}
Br(\mu\to e\gamma)=&\frac{\tau_\mu \alpha\; m_\mu^3}{16}
\frac{9 s_{2 \theta}^2}{(16\pi^2)^2}
\left\{
m^2_t \left| V_{ts}y^S_{21}y^R_{32} \right|^2
\left[ \frac{\frac{7}{6}+\frac{2}{3}\ln x^+_t}{M^2_+}  -\frac{\frac{7}{6}+\frac{2}{3}\ln x^-_t}{M^2_-}
 \right]^2
 \right.  \nonumber \\& \left. \hspace{3cm}
 +
 m^2_c \left| V_{cb}y^S_{32}y^R_{21} \right|^2
\left[ \frac{\frac{7}{6}+\frac{2}{3}\ln x^+_c}{M^2_+}  -\frac{\frac{7}{6}+\frac{2}{3}\ln x^-_c}{M^2_-}
 \right]^2
\right\}.
\end{align}

Finally, the combination of Eqs.~\eqref{eq:S1S3_t_c_aa}, \eqref{eq:S1S3_t_c_bb}, and \eqref{eq:S1S3_t_c_cc} yields
\begin{equation}
\label{eq:no_go_S1S3_t_c}
Br(\mu\to e\gamma)=
\frac{\tau_\mu \alpha\; m_\mu^3}{16}
\left(
\frac{\Delta a^2_e}{m^2_e}\;\frac{\widetilde{B}}{x^2} \frac{|V_{ts}|^2}{|V_{cs}|^2}+
\frac{\Delta a^2_\mu}{m^2_\mu}\;\frac{x^2}{\widetilde{B}} \frac{|V_{cb}|^2}{|V_{tb}|^2}
\right),
\end{equation}
where $x=y^R_{21}/y^R_{32}$ and
\begin{equation}
\widetilde{B}= \frac{m^2_t}{m^2_c} \frac{\left( \frac{\frac{7}{6}+\frac{2}{3}\ln x^+_t}{M^2_+}  -\frac{\frac{7}{6}+\frac{2}{3}\ln x^-_t}{M^2_-}
 \right)^2}{\left( \frac{\frac{7}{6}+\frac{2}{3}\ln x^+_c}{M^2_+}  -\frac{\frac{7}{6}+\frac{2}{3}\ln x^-_c}{M^2_-}
 \right)^2}.
\end{equation}
Even though the expression for $Br(\mu\to e\gamma)$ exhibits dependence on the scale of new physics the minimal attainable value $(Br(\mu\to e\gamma))^\mathrm{min}$ does not and is equal to the value already quoted in Eq.~\eqref{eq:no_go_S1_t_c_b}. Clearly, this particular $S_1 \& S_3$ scenario, when $y^S_{21}, y^S_{32}, y^R_{21}, y^R_{32} \neq 0$, fails to pass the $\mu\to e\gamma$ test since the fact that we work in the down-type quark mass-diagonal basis generates the CKM matrix induced coupling between the $Q=1/3$ leptoquark, a top quark and an electron.

The situation changes drastically if we work, with the same Yukawa ansatz, in the up-type quark mass-diagonal basis. This time around the $\mu\to e\gamma$ signature is completely absent since the switch to that basis would correspond to setting $|V_{ts}|=|V_{cb}|=0$ in Eq.~\eqref{eq:S1S3_t_c_cc}. We accordingly quote $(Br(\mu\to e\gamma))^\mathrm{min}=0$ in Table~\ref{tab:LQ_scenarios} for this particular $S_1 \& S_3$ scenario when the shifts in $(g-2)_\mu$ and $(g-2)_e$ are generated via the top quark and charm quark loops, respectively. Note that the $S_3^{4/3}$ state couples simultaneously to the muon-bottom quark and electron-bottom quark pairs and  could thus generate bottom quark mediated $\mu\to e\gamma$ loops. It turns out that these contributions to $\mu\to e\gamma$ are completely negligible within this particular Yukawa ansatz. What thus remains to be analysed is whether this scenario is also consistent with all other flavor and collider physics constraints. We address that question in what follows.

The relevant expressions for $\Delta a_e$ and $\Delta a_\mu$ are
\begin{align}
\label{eq:S1S3_t_c_aaa}
&\Delta a_e=-\frac{3m_em_c}{8\pi^2}\;y^R_{21}  y^S_{21}\;\left[ \frac{s_{2\theta}}{2 M^2_+} \left( \frac{7}{6}+\frac{2}{3}\ln x^+_c \right) -\frac{s_{2\theta}}{2 M^2_-}
\left( \frac{7}{6}+\frac{2}{3}\ln x^-_c \right) \right],
\\
\label{eq:S1S3_t_c_bbb}
&\Delta a_\mu=-\frac{3m_\mu m_t}{8\pi^2}\;y^R_{32}  y^S_{32}\;\left[ \frac{s_{2\theta}}{2 M^2_+} \left( \frac{7}{6}+\frac{2}{3}\ln x^+_t \right) -\frac{s_{2\theta}}{2 M^2_-}
\left( \frac{7}{6}+\frac{2}{3}\ln x^-_t \right) \right],
\end{align}
while the flavor constraints that we take into consideration are $K^0_L\to \pi^0\nu \nu$, $K^+\to \pi^+\nu \nu$, $B\to K^{{*}}\nu \nu$, $B_{b/s}\to \ell^+ \ell^{' -}$, $B^+\to \pi^+ \ell^+ \ell^{' -}$, $B^+\to K^+ \ell^+ \ell^{' -}$, $K^+\to \pi^+ \ell^+ \ell^{' -}$, $K^0_S\to \pi^0 \ell^+ \ell^{' -}$, $K^0_L\to \pi^0 \ell^+ \ell^{' -}$, $K^0_L\to \ell^+ \ell^{' -}$, $B^0_s-\overline{B}^0_s$, $B^0_d-\overline{B}^0_d$, $K^0-\overline{K}^0$, and
$Z\to \ell^+ \ell^-$. What we find is that this scenario is ruled out, at the $1\,\sigma$ level, by the interplay between the limits that originate from $K^0_L\to e^{\pm}\mu^{\mp}$, $Z\to \ell^+ \ell^-$, $\ell=e,\mu$, and the LHC data, as we show next.

Let us start with the $K^0_L\to e^{\pm}\mu^{\mp}$ constraint. The associated decay width is~\cite{Becirevic:2016zri}
\begin{align}
\Gamma =&\frac{1}{64\pi^3}\frac{\alpha^2 G^2_F f^2_K}{m^3_K}|V_{tb}V^*_{ts}|^2 \eta^{1/2}(m_K,m_e,m_\mu)
\nonumber \\& \times 
\left\{
\left[ m^2_K-(m_e+m_\mu)^2\right] (m_e-m_\mu)^2 |C_9|^2
+
\left[ m^2_K-(m_e-m_\mu)^2\right] (m_e+m_\mu)^2 |C_{10}|^2
\right\},
\end{align}
where
\begin{align}
&\eta^{1/2}(m_K,m_e,m_\mu)=  \sqrt{\left[ m^2_K-(m_e-m_\mu)^2\right] \left[ m^2_K-(m_e+m_\mu)^2\right] },
\\
&|C_9|^2=|C_{10}|^2=\frac{\pi^2 v^4}{\alpha^2 |V_{tb}V^*_{ts}|^2} \;\;|g|^2,
\\
&g= \frac{\left( V^T y^S \right)^*_{se}\left( V^T y^S \right)_{d\mu} +
\left( V^T y^S \right)^*_{de}\left( V^T y^S \right)_{s\mu}}{\sqrt{2}M^2_3}.
\end{align}
The branching ratio for the $K^0_L\to e^{\pm}\mu^{\mp}$ process, for $m_e=0$, is thus predicted to be
\begin{equation}
\label{eq:Kemu}
Br(K^0_L\to e^{\pm}\mu^{\mp}) =\frac{\tau_K f^2_K}{128\pi m^3_K} \frac{m^2_\mu (m^2_K - m^2_\mu)^2}{M^4_3}
\left|
\left( V^T y^S \right)^*_{se}\left( V^T y^S \right)_{d\mu} +
\left( V^T y^S \right)^*_{de}\left( V^T y^S \right)_{s\mu}
\right|^2.
\end{equation}
The current experimental limit $Br(K^0_L\to e^{\pm}\mu^{\mp})^\mathrm{exp} < 4.7 \times 10^{-12}$, combined with $f_K=161$\,MeV, $m_K=497.116$\,MeV, and $\tau_K=7.77632\times 10^{16}$\,GeV$^{-1}$, then yields the following limit on the Yukawa couplings of $S_3$ leptoquark to the left-chiral leptons
\begin{align}
|y^S_{21}y^S_{32}| < 5.144\times 10^{-4} \left( \frac{M_3}{1\,\mathrm{TeV}} \right)^2.  
\end{align}

Simply put, the $K^0_L\to e^{\pm}\mu^{\mp}$ and the current LHC data on the leptoquark searches imply that the product of the left-chiral couplings should be rather small. The need to generate substantial shifts of $(g-2)_e$ and $(g-2)_\mu$, as given in Eqs.~\eqref{eq:S1S3_t_c_aaa} and \eqref{eq:S1S3_t_c_bbb}, consequently requires the $S_1$ couplings to the right-chiral leptons to be large. But, the largeness of these couplings is in tension with the $Z\to \ell^+ \ell^-$ data, as we show next.

Namely, following Ref.~\cite{Arnan:2019olv}, we parametrise the relevant interactions of the $Z$ boson with
\begin{align}
\delta \mathcal{L}^{Z\to \ell \ell^{\prime}}_{eff}=\frac{g}{\cos \theta_W} \sum_{f,i,j} \overline{f}_i \gamma^{\mu} \left( g^{ij}_{L}P_L + g^{ij}_{R}P_R \right) f_j Z_{\mu},   
\end{align}
where $g^{ij}_{L}$ and $g^{ij}_{R}$ measure the strength of interaction between the $Z$ boson and the left- and right-chiral fermions, respectively. Since $S_1$ leptoquark, in our case, couples purely to the right-chiral leptons the current experimental limits on $Z\to e^+ e^-$ and $Z\to \mu^+\mu^-$~\cite{ALEPH:2005ab} provide the following constraints, at the $1\,\sigma$ level, on the shift of $g^{ij}_{R}$ with respect to the SM value: 
\begin{align}
&Re[\delta g^{ee}_R]\leq 2.9\times 10^{-4},\;\;  Re[\delta g^{\mu\mu}_R]\leq 1.3\times 10^{-3}.
\end{align}

The exact expressions for $\delta g^{\ell\ell}_R$, $\ell=e,\mu$, are
\begin{align}
\delta g^{ee}_R&= \frac{1}{16\pi^2}
x^+_Z w^+_{ce}w^{+*}_{ce} \left[ g_{u_R} \left( \ln x^+_Z -i\pi -\frac{1}{6} \right) +\frac{g_{\ell_R}}{6} \right]
\nonumber \\& + \frac{1}{16\pi^2}
x^-_Z w^-_{ce}w^{-*}_{ce} \left[ g_{u_R} \left( \ln x^-_Z -i\pi -\frac{1}{6} \right) +\frac{g_{\ell_R}}{6} \right],\\
\delta g^{\mu\mu}_R&= \frac{3}{16\pi^2}
w^+_{t\mu}w^{+*}_{t\mu} \left[ (g_{u_R}-g_{u_L}) \frac{x^+_t\left( x^+_t-1-\ln x^+_t \right)}{(x^+_t-1)^2} +\frac{x^+_Z}{12} F^R(x^+_t)\right]
\nonumber \\& + \frac{3}{16\pi^2}
w^-_{t\mu}w^{-*}_{t\mu} \left[ (g_{u_R}-g_{u_L}) \frac{x^-_t\left( x^-_t-1-\ln x^-_t \right)}{(x^-_t-1)^2} +\frac{x^-_Z}{12} F^R(x^-_t)\right],
\end{align}
where $w^+_{q\ell}=\left( c_\theta y^R \right)_{q\ell}$, $w^-_{q\ell}=\left( s_\theta y^R \right)_{q\ell}$, and $x^{\pm}_Z=m_Z^2/M_{\pm}^2$, while the function $F^R(x)$ is given in Ref.~\cite{Arnan:2019olv}.

\begin{figure}
\centering
     \includegraphics[width=.45\textwidth]{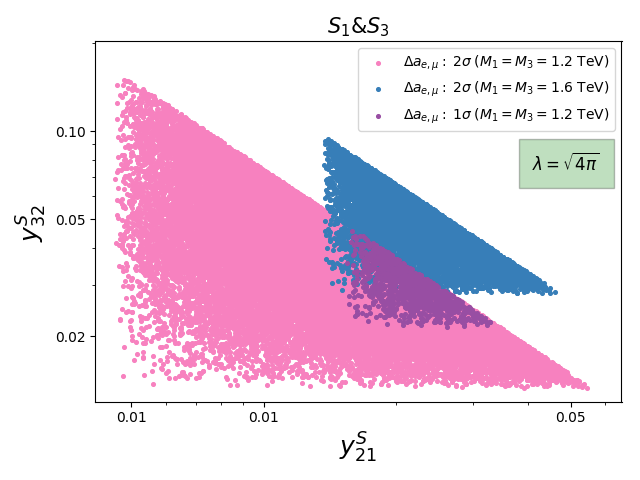}
     \includegraphics[width=.45\textwidth]{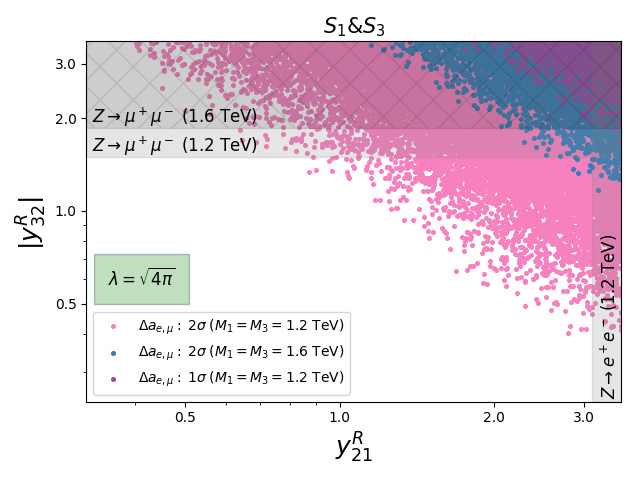}
     \caption{The available parameter space for the left-chiral couplings of $S_3$ (left-hand panel) and the right-chiral couplings of $S_1$ (right-hand panel) with respect to the $K^0_L\to e^{\pm}\mu^{\mp}$ and $Z\to \ell^+ \ell^-$ constraints for two different inputs for the leptoquark masses. The $(g-2)_{e,\mu}$ discrepancies are addressed at the $1\,\sigma$ or $2\,\sigma$ levels, as indicated.}
     \label{fig1}
\end{figure}

For example, if take $M_1=M_3=1$\,TeV, and $\lambda=3$ the inferred $1\,\sigma$ limits are $y^R_{21} \leq 2.67$ and $y^R_{32} \leq 1.33$. These limits, once again, would turn out to be rather constraining factor for the $S_1\&S_3$ scenario. We accordingly present in Fig.~\ref{fig1} the interplay between the $K^0_L\to e^{\pm}\mu^{\mp}$ and $Z\to \ell^+ \ell^-$ constraints. The left-hand panel of Fig.~\ref{fig1} shows the available parameter space for the left-chiral couplings of $S_3$ leptoquark if we demand that the $(g-2)_{e,\mu}$ discrepancies are addressed at the $1\,\sigma$ or $2\,\sigma$ levels, as indicated, for two different leptoquark mass inputs. The choice $M_1=M_3=1.2$\,TeV is not realistic in view of the latest LHC data on the leptoquark searches but we choose to show it in order to illustrate the mass dependence of our results. Note that in the $S_1\&S_3$ scenario that we consider there are four physical leptoquarks, three of which can decay into charged leptons. The two $Q=1/3$ states decay into $t\mu$ and $ce$ pairs, whereas the $Q=4/3$ state can primarily decay into $b\mu$, $be$, and $se$ pairs. We find that the most relevant bounds on the leptoquark masses come from the direct searches for the scalar leptoquark pair production at LHC via the final states that feature $t\mu$ or $b\mu$ pairs. For example, the current limits on the mass $M_S$ of scalar leptoquark $S$ are $M_S \gtrsim 1470$\,GeV and $M_S \gtrsim 1400$\,GeV for $\mathcal{B}(S\to t\mu)=1$~\cite{ATLAS:2020qoc} and $\mathcal{B}(S\to b\mu)=1$~\cite{Diaz:2017lit}, respectively. Since we have multiple leptoquarks that decay into the same final state we use the approach advocated in Ref.~\cite{Diaz:2017lit} to deduce that the conservative mass limit on scalar leptoquarks in the $S_1\&S_3$ scenario is $M_S \gtrsim 1.6$\,TeV. Note that the actual branching ratios for the leptoquark decay channels vary as one moves within the $(g-2)_{e,\mu}$ preferred Yukawa parameter space. Hence the need for the conservative limit on the leptoquark masses. The $M_1=M_3=1.6$\,TeV input that we use is thus realistic with respect to the LHC data but it is viable only at the $2\,\sigma$ level when it comes to the flavor physics constraints. We take, for both mass inputs, the maximal value for the mixing parameter $\lambda$, as allowed by the perturbativity arguments, to maximise the impact of the $S_1\&S_3$ scenario on the $(g-2)_{e,\mu}$ discrepancies. The right-hand panel of Fig.~\ref{fig1} shows the corresponding parameter space for the right-chiral couplings of $S_1$ leptoquark. Clearly, there is only a small part of parameter space that passes the $Z \to \mu^+ \mu^-$ constraint at the $2\sigma$ level for the $M_1=M_3=1.6$\,TeV input. Note that the perturbativity limit on $y_{21}^R$, i.e., $|y_{21}^R| \leq \sqrt{4 \pi}$, is stronger than the $Z \to e^+ e^-$ constraint in that particular leptoquark mass scenario.

To conclude, the interplay between $K^0_L\to e^{\pm}\mu^{\mp}$ and $Z\to \ell^+ \ell^-$ practically precludes combined explanation of the $(g-2)_{e,\mu}$ discrepancies at the $1\sigma$ level for the realistic input for the leptoquark mass spectrum. We have performed our analysis assuming that all the Yukawa couplings as well as the CKM matrix elements are real, for simplicity. Also, the $S_1\&S_3$ scenario might leave an imprint on the $h \to \mu^+ \mu^-$ process, as discussed recently in Ref.~\cite{Crivellin:2020tsz}, but this type of analysis also depends on additional terms in the Lagrangian that are not featured in our manuscript and we accordingly opt not to perform it.

\subsubsection{\texorpdfstring{$R_2 \& \widetilde{R}_2$}{R2 \& R2 tilde} with the top and bottom quark loops}

We consider the $R_2 \& \widetilde{R}_2$ scenario when $(g-2)_\mu$ is addressed via the top quark loops and with Yukawa couplings of $R^{5/3}_2$ state while the $(g-2)_e$ discrepancy, on the other hand, is addressed with the bottom quark loops and with Yukawa couplings that are associated with the mixture of $R^{2/3}_2$ and $\widetilde{R}^{2/3}$ states. The relevant parts of the Lagrangian that have not been featured in the preceding sections are
\begin{align}
\label{eq:R2t_aa}
\mathcal{L} \supset -\widetilde{y}_{ij}^L \overline{d}_{R i} \widetilde{R}_2 i\sigma_2 L_{L_{j}}  - \lambda \, \big{(}R_2^\dagger H\big{)}\big{(}\widetilde{R}_2^T i \sigma_2 H\big{)} + \text{h.c.},
\end{align}
where $\overline{d}_{R i}$ are the right-handed down-type quarks and $\lambda$ is a dimensionless coupling.

The Yukawa couplings of the $\widetilde{R}_2$ charge eigenstate components, in the mass eigenstate basis for the SM fermions, are
\begin{eqnarray}
\label{eq:R2t}
\mathcal{L}_{\widetilde{R}_2} &=& - (D_R^\dagger \widetilde{y}^L E_L)_{i j} \overline{d}_{R i} \ell_{L j} \widetilde{R}_2^{2/3} + (D_R^\dagger \widetilde{y}^L N_L)_{i j} \, \overline{d}_{Ri} \nu_{L j} \widetilde{R}_2^{-1/3} + \text{h.c.}.
\end{eqnarray}
We judiciously switch on $\widetilde{y}^L_{31}$ in Eq.~\eqref{eq:R2t_aa} and $y^L_{32}$, $y^R_{31}$, and $y^R_{32}$ in Eq.~\eqref{eq:R2} to generate the loops of interest in the down-type quark mass-diagonal basis. Note that $R_2^{2/3}$ needs to couple to the right-chiral electron and a bottom quark for the $R_2 \& \widetilde{R}_2$ scenario to work. Consequentially, three Yukawa couplings are associated with the $R_2$ and only one Yukawa is associated with $\widetilde{R}_2$. The mixed states of electric charge $Q=2/3$ are
\begin{align}
&R^{2/3}_2=c_\theta R_- -s_\theta R_+\\
&\widetilde{R}^{2/3}_2=s_\theta R_- +c_\theta R_+,
\end{align}
with the mixing angle $\theta$ defined via 
\begin{equation}
\tan 2\theta = \frac{\lambda v^2}{\widetilde{M}^2 - M^2},
\end{equation}
and
\begin{equation}
M^2_{\pm}=\frac{\widetilde{M}^2+M^2}{2}\pm \frac{1}{2} 
\sqrt{ (\widetilde{M}^2-M^2)^2+\lambda^2v^4 },
\end{equation}
where the new physics mass eigenstates, after the mixing takes place, are $R_2^{5/3}$, $R_{\pm}$, and $\widetilde{R}_2^{-1/3}$ with $M=M_{R_2^{5/3}}$, $M_{\pm}=M_{R_{\pm}}$, and $\widetilde{M}=\widetilde{M}_{\widetilde{R}_2^{-1/3}}$.   

For real Yukawa couplings we obtain that
\begin{align}
\label{eq:R2R2t_a}
&\Delta a_e=-\frac{m_em_b}{32 \pi^2}s_{2 \theta}\; y^R_{31} \widetilde{y}^L_{31} \left( \frac{5+2\ln x^+_b}{M^2_+} -  \frac{5+2\ln x^-_b}{M^2_-} 
\right),
\\
\label{eq:R2R2t_b}
&\Delta a_\mu=-\frac{m_\mu m_t}{16\pi^2}\; y^R_{32} y^L_{32} \left( \frac{1+4\ln x_t}{M^2}
\right),
\\
\label{eq:R2R2t_c}
&Br(\mu\to e \gamma)= \frac{\tau_\mu \alpha m^3_\mu m^2_t}{4096 \pi^4}\times 
\nonumber \\& 
\left[
\left| y^R_{31}y^L_{32} \left( \frac{1+4\ln x_t}{M^2} \right)   \right|^2
+ \left( \frac{s_{2 \theta} m_b}{2 m_t} \right)^2
\left| \widetilde{y}^L_{31}y^R_{32} \left( \frac{5+2\ln x^+_b}{M^2_+} -  \frac{5+2\ln x^-_b}{M^2_-} 
\right)   \right|^2
\right],
\end{align}
where $x^\pm_b=m_b^2/M_{\pm}^2$ and $m_b$ is the bottom quark mass.

If we introduce $x=y^R_{31}/y^R_{32}$ and rewrite $Br(\mu\to e\gamma)$ in Eq.~\eqref{eq:R2R2t_c} using Eqs.~\eqref{eq:R2R2t_a} and~\eqref{eq:R2R2t_b} we find that
\begin{equation}
Br(\mu\to e\gamma)=
\frac{\tau_\mu \alpha\; m_\mu^3}{16}
\left(
\frac{\Delta a^2_e}{m^2_e}\;\frac{1}{x^2}+
\frac{\Delta a^2_\mu}{m^2_\mu}\;x^2\right).
\end{equation}
This expression yields the same minimal value for $Br(\mu\to e\gamma)$ as given in Eq.~\eqref{eq:no_go_S1_t_t_b} that is eigth to nine orders of magnitude above the current experimental limit, which demonstrates that the limit on $\mu\to e\gamma$ is an unsurmountable obstacle for this particular scenario.

Note, however, that it is possible to significantly suppress $\mu\to e\gamma$ signature within the $R_2 \& \widetilde{R}_2$ scenario with somewhat special Yukawa ansatz. Namely, if we work in the up-type quark mass-diagonal basis and we switch on $\widetilde{y}^L_{31}$ in Eq.~\eqref{eq:R2t_aa} and $y^L_{32}$, $y^R_{21}$, and $y^R_{32}$ in Eq.~\eqref{eq:R2} we can generate the $(g-2)_{e,\mu}$ loops of interest but with the $\mu\to e\gamma$ signature being generated only via the bottom quark mediation. The relevant expressions for $\Delta a_e$, $\Delta a_\mu$, and $Br(\mu\to e \gamma)$ are
\begin{align}
\label{eq:R2R2t_aa}
&\Delta a_e=-\frac{m_em_b}{32 \pi^2}s_{2 \theta}\; V_{cb} y^R_{21} \widetilde{y}^L_{31} \left( \frac{5+2\ln x^+_b}{M^2_+} -  \frac{5+2\ln x^-_b}{M^2_-} 
\right),
\\
\label{eq:R2R2t_bb}
&\Delta a_\mu=-\frac{m_\mu m_t}{16\pi^2}\; y^R_{32} y^L_{32} \left( \frac{1+4\ln x_t}{M^2}
\right),
\\
\label{eq:R2R2t_cc}
&Br(\mu\to e \gamma)= \frac{\tau_\mu \alpha m^3_\mu }{4096 \pi^4} \left( \frac{s_{2 \theta} m_b}{2} \right)^2
\left| V_{tb} \widetilde{y}^L_{31}y^R_{32} \left( \frac{5+2\ln x^+_b}{M^2_+} -  \frac{5+2\ln x^-_b}{M^2_-} 
\right)   \right|^2.
\end{align}

To obtain $(Br\left(\mu\to e\gamma\right))^\mathrm{min}$ we first combine Eqs.~\eqref{eq:R2R2t_aa} and \eqref{eq:R2R2t_cc}. This yields
\begin{equation}
\label{eq:R2R2t_dd}
Br(\mu\to e \gamma)= \frac{\tau_\mu \alpha m^3_\mu }{16 m_e^2} \frac{|V_{tb}|^2}{|V_{cb}|^2} \frac{|y^R_{32}|^2}{|y^R_{21}|^2} |\Delta a_e|^2,
\end{equation}
where the minimum for $Br(\mu\to e \gamma)$ would be obtained if we maximize $y^R_{21}$ and minimize $y^R_{32}$. If we accordingly set $y^R_{21}=y^L_{32}=\sqrt{4 \pi}$ and insert Eq.~\eqref{eq:R2R2t_bb} into Eq.~\eqref{eq:R2R2t_dd}, we finally get that
\begin{equation}
\label{eq:R2R2t_dd}
(Br\left(\mu\to e\gamma\right))^\mathrm{min}= \frac{\tau_\mu \alpha\; m_\mu^3 \pi^2}{m_e^2 m_\mu^2 m_t^2} \frac{|V_{tb}|^2}{|V_{cb}|^2}
\frac{|\Delta a_e \Delta a_\mu|^2}{(1+4 \ln x_t)^2} M^4,
\end{equation}
where $M$, again, is the mass of $R_2^{5/3}$. For example, if we take $M = 1.6$\,TeV we obtain $(Br\left(\mu\to e\gamma\right))^\mathrm{min}$ that exceeds current experimental limit by three orders of magnitude. Note that Eq.~\eqref{eq:R2R2t_dd} is valid roughly up to $M <160$\,TeV since we demand that $y^R_{32}$ remains perturbative at all times. It is $(Br\left(\mu\to e\gamma\right))^\mathrm{min}$, as given in Eq.~\eqref{eq:R2R2t_dd}, that we quote in Table~\ref{tab:LQ_scenarios}.

\section{Conclusions}
\label{sec:conclusion}

We investigate all possible ways to simultaneously address discrepancies between the observed values of the electron and muon anomalous magnetic moments and the SM theoretical predictions with the new physics scenarios that introduce one or, at most, two scalar leptoquarks. We provide classification of these scenarios in terms of their ability to satisfy the existing limits on the branching ratio for the $\mu \to e \gamma$ process. In order to be of the correct strength the chirality enhanced $(g-2)_e$ loops could be due to the top quark, charm quark, or bottom quark propagation, while the $(g-2)_\mu$ loops should be generated solely through the top quark propagation. The simultaneous explanation of the discrepancies, on the other hand, coupled with the current experimental data, requires that the $(g-2)_e$ loops are exclusively due to the charm quark propagation. If the $(g-2)_e$ loops are due to the top (bottom) quark propagation the predicted minimal value for the $\mu\to e\gamma$ branching ratio exceeds the associated experimental limit by approximately nine (three) orders of magnitude. The scenarios we consider require at least four Yukawa couplings to be switched on in order to generate the aforementioned loops, where one pair feeds into the electron anomalous magnetic moment and the other pair into the muon one. 

There are only three particular scenarios that can pass the $\mu \to e \gamma$ test and create large enough impact on the $(g-2)_{e,\mu}$ discrepancies when the new physics is based on the SM fermion content. These are the $S_1$, $R_2$, and $S_1 \& S_3$ scenarios, where the first two are already known to be phenomenologically viable candidates with respect to all other flavor and collider data constraints. We show that the third scenario, where the right-chiral couplings to leptons are due to $S_1$, the left-chiral couplings to leptons are due to $S_3$, and the two leptoquarks mix through the SM Higgs field, cannot simultaneously address the $(g-2)_{e,\mu}$ discrepancies at the $1\sigma$ level due to an interplay between $K_L^0 \to e^\pm \mu^\mp$, $Z \to e^+ e^-$, and $Z \to \mu^+ \mu^-$ data despite the ability of that scenario to avoid the $\mu \to e \gamma$ limit.

\section*{Acknowledgments}
\label{sec:acknowledgments}
We thank O.\ Sumensari for constructive discussions. S.F.\ acknowledges the financial support from the Slovenian Research Agency (research core funding No.\ P1-0035).

\bibliographystyle{bibstyle}
\bibliography{references}
\end{document}